\documentclass[letterpaper,12pt]{article}
\pdfoutput=1

\usepackage{jheppub}
\usepackage{amsmath}
\usepackage{graphicx}
\usepackage{subfig}
\usepackage{cancel}
\usepackage[dvipsnames]{xcolor}
\usepackage{color}
\usepackage{mathrsfs}

\usepackage{amssymb}
\usepackage{epstopdf}
\newpage

\title{Restricted Maximin surfaces and HRT in generic black hole spacetimes}

\author[a]{Donald Marolf,}
\author[b]{Aron C. Wall,}
\author[a]{Zhencheng Wang}

\affiliation[a]{Department of Physics, University of California, Santa Barbara, CA 93106, USA}
\affiliation[b]{Stanford Institute for Theoretical Physics, 382 Via Pueblo, Stanford University, Stanford, CA 94305, USA}

\emailAdd{marolf@physics.ucsb.edu}
\emailAdd{aroncwall@gmail.com}
\emailAdd{zhencheng@physics.ucsb.edu}

\abstract{The AdS/CFT understanding of CFT entanglement is based on HRT surfaces in the dual bulk spacetime.   While such surfaces need not exist in sufficiently general spacetimes,  the maximin construction demonstrates that they can be found in any smooth asymptotically locally AdS spacetime without horizons or with only Kasner-like singularities.   In this work, we introduce restricted maximin surfaces anchored to a particular boundary Cauchy slice $C_\partial$. We show that the result agrees with the original unrestricted maximin prescription when the restricted maximin surface lies in a smooth region of spacetime.  We then use this construction to extend the existence theorem for HRT surfaces to generic charged or spinning AdS black holes whose mass-inflation singularities are not Kasner-like. We also discuss related issues in time-independent charged wormholes.
}

\begin{document}
\maketitle

\section{Introduction}
\label{sec:Introduction}
As is by now well established \cite{Lewkowycz:2013nqa,Dong:2016hjy}, in AdS/CFT the Ryu-Takayangi \cite{Ryu:2006bv,Ryu:2006ef} and Hubeny-Rangamani-Takayanagi (HRT) \cite{Hubeny:2007xt} prescriptions generally describe the von Neumann entropy of CFT regions $A$ in terms of the area of an appropriate bulk surface.  In particular,
\begin{equation}
S _ { A } =   \frac { \operatorname { Area } [ \operatorname{ext} ( A ) ] } { 4  G } ,
\end{equation}
where $\operatorname{ext} ( A ) $ is the smallest extremal surface satisfying $\partial(\operatorname{ext} ( A ))=\partial A$ and with $\operatorname{ext} ( A )$ homologous to $A$. When there is more than one such surface with minimal area, the HRT surface is ambiguous.  Such situations arise at HRT phase transitions, when the HRT surface jumps discontinuously as one varies the region $A$.

Now, there are spacetimes in which HRT surfaces fail to exist or where those that do exist do not correctly compute the von Neumann entropy \cite{Fischetti:2014uxa}.  However, known spacetimes $M_0$ with the latter issue are $\lambda \rightarrow 0$ limits of spacetimes $M_\lambda$ in which the HRT prescription succeeds, but where the correct (smallest) extremal surface recedes to the future or past singularity as $\lambda \rightarrow 0$.  Similarly, known spacetimes $M'_0$ where extremal surfaces fail to exist are again $\lambda \rightarrow 0$ limits of spacetimes $M_\lambda$ where
HRT succeeds but in which {\it all} extremal surfaces recede in this way.

One thus expects that HRT surfaces do in fact correctly compute the entropy in contexts such recessions are forbidden; i.e., where extremal surfaces are guaranteed to exist as surfaces in smooth regions of the bulk.  The maximin construction of \cite{Wall:2012uf} shows this to be the case in asymptotically locally-AdS (AlAdS) spacetimes without horizons or where the future and past boundaries consist only of Kasner-like singularities\footnote{In contrast, the examples of \cite{Fischetti:2014uxa} contain smooth de Sitter-like pieces of future or past infinity as well as special non-Kasner-like singular points where the smooth parts of future/past infinity meet otherwise-Kasner-like singularities.}. Ref. \cite{Wall:2012uf} also shows in this context that HRT surfaces satisfy strong subadditivity.

\begin{figure}
\begin{center}
  \includegraphics[width=0.2\textwidth]{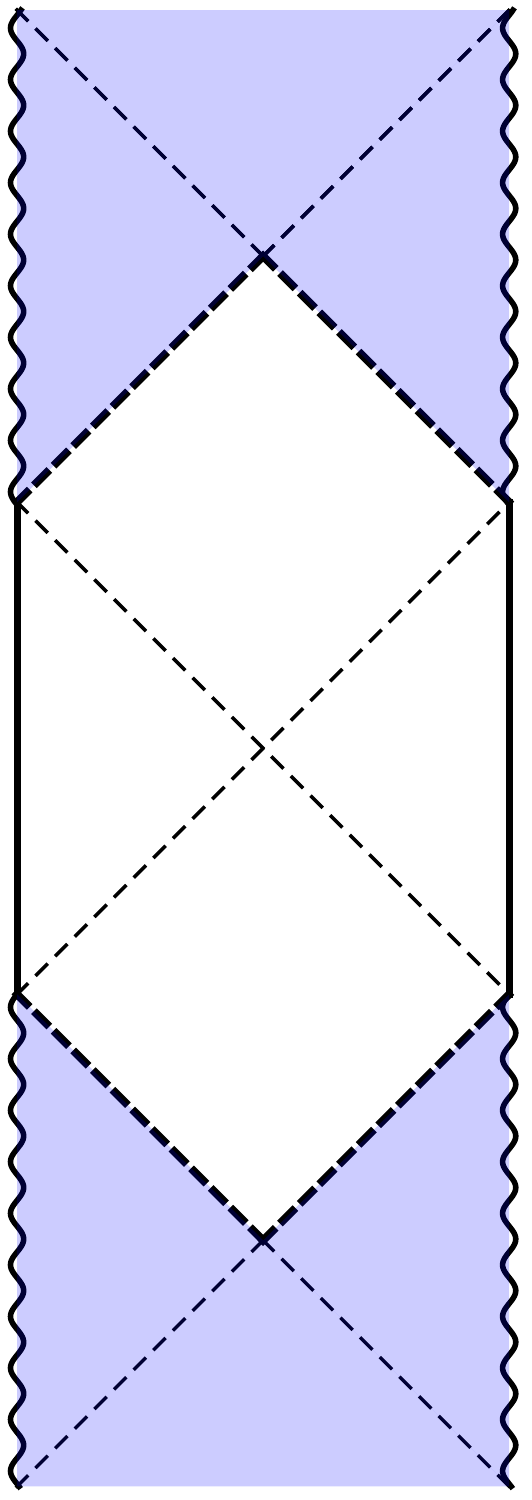}
  \caption{The maximal analytic extension of the AdS-Reissner-Nordstr\"om black hole.  for our study in Section \ref{sec:existence} we truncate it to the AdS-hyperbolic unshaded region between the past and future (AdS-) Cauchy horizons (heavy dashed lines).}
  \label{fig:AdSRN}
 \end{center}
\end{figure}

\begin{figure}
\begin{minipage}[t]{0.5\textwidth}
\centering
  \includegraphics[width=0.3\textwidth]{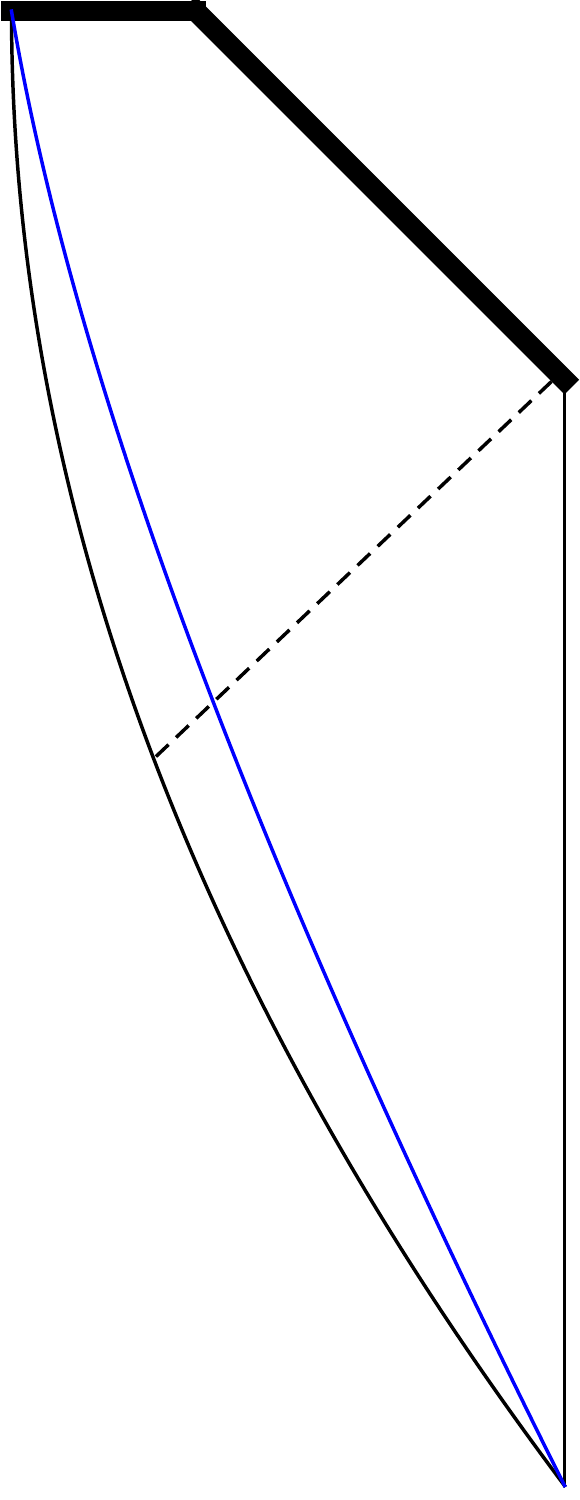}
  \end{minipage}
  \begin{minipage}[t]{0.5\textwidth}
\centering
  \includegraphics[width=0.4\textwidth]{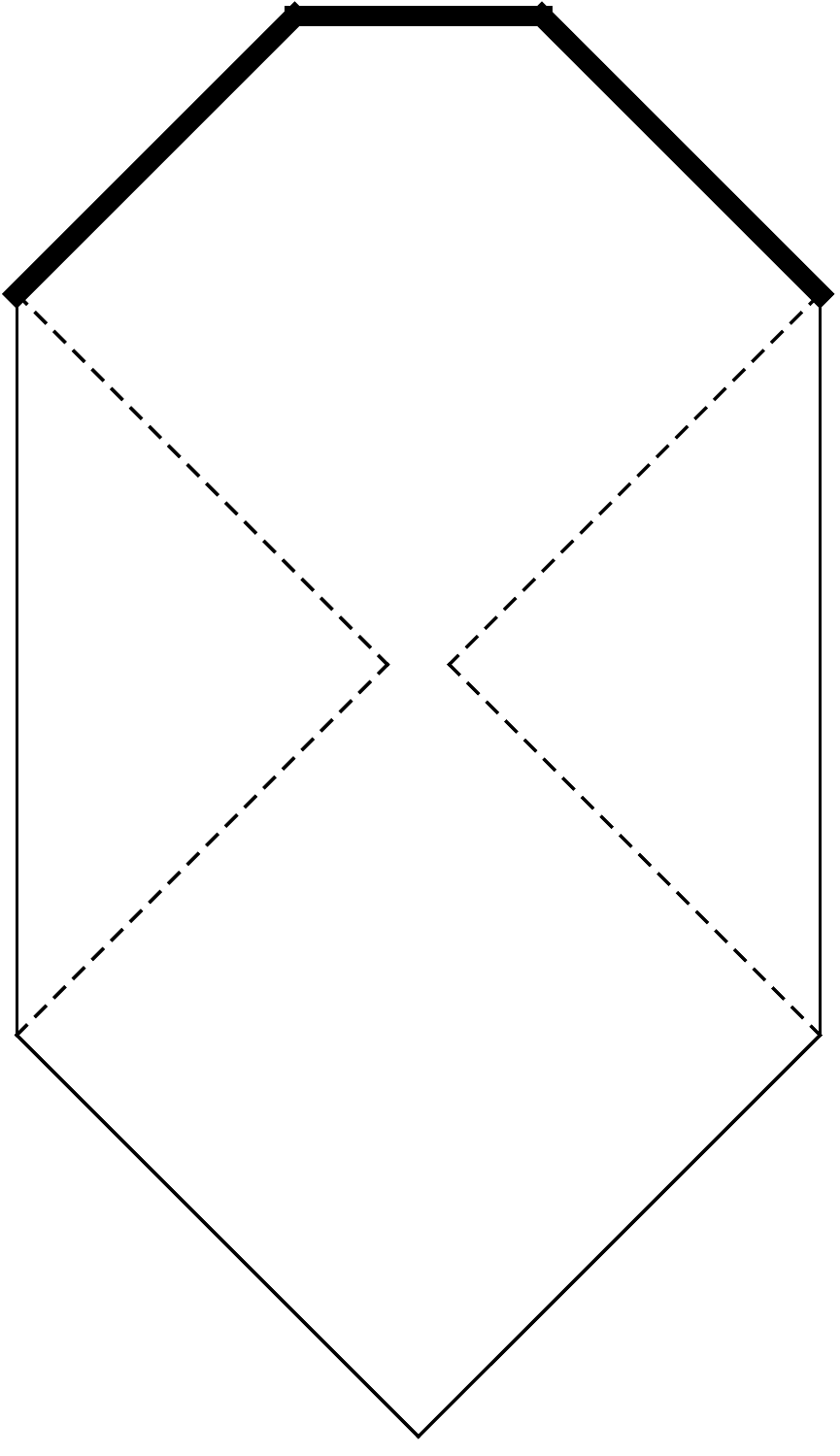}
  \end{minipage}
  \caption{Perturbed one-sided (left) and two-sided (right) AdS-RN black holes. The null parts are mass-inflation singularities.
  A spacelike piece of the singularity forms whenever caustics arise on a null singularity. Such caustics always arise in the one-sided case, and also occur for strong enough perturbations (as shown here) in the two-sided case.  The resulting spacelike singularities should be Kasner-like, as can be seen from the fact that the region between the inner- and outer-horizons in figure \ref{fig:AdSRN} admits a foliation by spatially homogenous slices that, when subjected to correspondingly homogeneous perturbations, becomes precisely an AdS-Kasner solution.  Sufficiently close to a curvature singularity, one should be able to treat any solution as approximately homogeneous, so the spacelike part of the singularity should again be Kasner-like.   In the left panel, the black hole is formed by a collapsing shell (in blue).}
  \label{fig:massinf}
\end{figure}

However, the full array of possible spacetimes have not yet been explored.  Of particular interest are charged or rotating black holes.  As is well known, stationary such black holes generally contain Cauchy horizons (see figure \ref{fig:AdSRN} for the AdS-Reissner-Nordstr\"om [AdS-RN] case). But this structure is unstable, and perturbations transform the Cauchy horizons into null mass-inflation singularities which are {\it not} Kasner-like \cite{Poisson:1989zz,Ori:1991zz,Ori:1992zz,Burko:1997zy,Burko:1997fc,
Dafermos:2003wr,Bhattacharjee:2016zof}; see figure \ref{fig:massinf}.
As discussed in the above references, generic black holes are believed to contain singularities of this type.   We show below that HRT surfaces exist in such spacetimes as well.

Our method of proof extends the maximin arguments of \cite{Wall:2012uf}. As defined in  \cite{Wall:2012uf}, a maximin surface is a codimension-2 surface anchored to $\partial A$ and satisfying the homology constraint, and minimizing area within some bulk Cauchy surface $\Sigma \supset A$, but which is also maximal among such minimal surfaces with respect to variations of $\Sigma$. In particular, the intersection of $\Sigma$ with the AlAdS boundary is allowed to vary so long as it still contains $\partial A$.  Below, we consider restricted maximin surfaces -- defined by bulk Cauchy surfaces $\Sigma$ that intersect the AlAdS boundary on a {\it fixed} boundary Cauchy surface $C_\partial$ -- and show that they must agree with with HRT surfaces (and thus with unrestricted maximin surfaces) when they lie in a smooth region of the spacetime.  In particular, since any Cauchy surface $\Sigma$ is achronal, restricted maximin surfaces must be achronally related to some $C_\partial$.  They are thus forbidden from reaching the null singularities in figure \ref{fig:massinf} and must lie in the smooth interior of the bulk spacetime as desired.

We begin by introducing restricted maximin surfaces in section \ref{sec:restricted}
and showing their equivalence to HRT surfaces when they lie in a smooth region of spacetime.  Existence of HRT surfaces in (perturbed) AdS-RN-like spacetimes then follows immediately, and more generally in spacetimes where boundary-anchored bulk Cauchy surfaces can reach a future boundary only at Kasner-like singularities. Section \ref{sec:discussion} concludes with a brief discussion of possible extensions to spacetimes with more complicated null singularities.

\section{Restricted maximin surfaces}
\label{sec:restricted}

This section will discuss restricted maximin surfaces.
In a different context, a maximin construction that fixes the entire boundary of a (in that case partial) Cauchy surface was also used in \cite{Sanches:2016sxy}.
Here and below we assume i) the null curvature condition (NCC): $R_{ab} k^ak^b \ge0$ at each point for every null vector $k^a$, ii) the generic condition \cite{Wald:1984rg}, which requires at least some positive focusing along each segment of any null geodesic\footnote{In fact, for our purposes it suffices for the spacetime to be a limit of spacetimes in which the generic condition holds, where the amount of focusing can vanish in the limit.  This will be the case in examples like exact AdS-RN discussed below in which the generic condition does not hold.}, and iii) AdS-hyperbolicity in the sense of \cite{Wall:2012uf}. We choose an achronal codimension-1 surface $A$ in the AlAdS boundary $\partial M$ to define the boundary region whose entropy we wish to study.  The boundary of $A$ is denoted $\partial A$.  Our restricted maximin surfaces are then defined via the following two-step procedure.

{\bf Definition 1}: For a chosen Cauchy surface $C_\partial$ of $\partial M$ with satisfies $A \subset C_\partial$, on any complete bulk Cauchy surface $\Sigma$ with $\Sigma \cap \partial M = C_\partial$ let $\min (A,\Sigma,C_\partial)$ denote the minimal-area codimension-2 surface anchored to $\partial A$ and homologous to $A$ within $\Sigma$ (i.e., such that there is a region $R$ of $\Sigma$ for which $\partial R = A \cup \min (A,\Sigma,C_\partial)$).

If there are multiple minimal area surfaces on $\Sigma$, then $\min(A, \Sigma,C_\partial)$ can refer to any of them.

{\bf Definition 2}: The restricted maximin surface $M_R (A,C_\partial)$ is defined as the $\min (A,\Sigma,C_\partial)$ whose area is maximal with respect to variations of $\Sigma$ that preserve $C_\partial$.  We use $\Sigma_{M_R (A,C_\partial)}$ to denote a Cauchy surface on which $M_R (A,C_\partial)$ is minimal.

In the case where there are multiple such surfaces, let $M_R (A,C_\partial)$ denote any such surface that is stable in the following sense: When $\Sigma$ is deformed infinitesimally to any nearby slice $\Sigma^\prime$ (still containing $C_\partial$), the new $\Sigma'$ still contains a {\it locally}-minimal surface $M'_R (A,C_\partial)$ on $\Sigma^\prime$ close to $M_R (A,C_\partial)$ which has no greater area, i.e. $\operatorname { Area }[M_R' (A,C_\partial)] \leq \operatorname { Area }[M_R (A,C_\partial)]$\footnote{This definition of stability fixes certain difficulties with the definition given in \cite{Wall:2012uf}.  A similary improved version of \cite{Wall:2012uf} will appear soon. }.

Below, we follow \cite{Wall:2012uf} in assuming that the stability criterion can be satisfied.  When $\Sigma_{M_R (A,C_\partial)}$ is both spacelike and smooth, this follows by the technical argument in section 3.5 of \cite{Wall:2012uf}.  But it remains an assumption more generally.  Existence of $M_R (A,C_\partial)$ then follows as in section 3.4 of
\cite{Wall:2012uf} so long as boundary-anchored Cauchy surfaces can future or past boundaries only at Kasner-like singularities.  In particular, the space of boundary-anchored achronal slices is compact in the same sense as the space of achronal slices anchored only to $\partial A$.

\subsection{Equivalence of HRT surfaces and restricted maximin surfaces in smooth regions of spacetime}
\label{sec:equiv}

We now show that the restricted maximin surface $M_R (A,C_\partial)$ is an HRT surface for every choice of $C_\partial$ that contains $A$ so long as $M_R (A,C_\partial)$  lies in a smooth region of the bulk spacetime.  The argument follows that given in \cite{Wall:2012uf} for the original unrestricted maximin surfaces.

We first show that $M_R (A,C_\partial)$ extremizes the area with respect to all variations that preserve $\partial A$. We begin with the case where $\Sigma_{M_R (A,C_\partial)}$ has continuous first derivative.  For every point on a restricted maximin surface $M_R (A,C_\partial)$, there are two independent directions that are normal to $M_R (A,C_\partial)$. The area is minimal with respect to variations on $\Sigma_{M_R (A,C_\partial)}$, and maximal with respect to variations normal to this surface.  The corresponding first order variations of the area vanish.  Linearity of first order variations then implies the area to be stationary under all deformations that preserve $\partial A$; i.e., the surface is extremal as desired.

If instead the first derivative of $\Sigma_{M_R (A,C_\partial)}$ jumps discontinuously, the surface $M_R (A,C_\partial)$ must still be extremal. The argument is identical to that of Theorem 15(b) in \cite{Wall:2012uf}.

We now show that $M_R (A,C_\partial)$ is the (properly anchored) extremal surface with least area, and thus an HRT surface.  The argument uses the notion introduced in \cite{Wall:2012uf} of the `representative' of any extremal surface $x(A)$ on a Cauchy surface $\Sigma$. The representative $\tilde x_\Sigma (A)$  is defined by observing that $x(A)$ splits some Cauchy surface into two pieces, which we arbitrarily label as $\Sigma_1$ and $\Sigma_2$.  When the new Cauchy surface $\Sigma$ lies to the future of $\Sigma_1$, this representative may be taken to be the intersection of $\Sigma$ with the boundary of the future of $\Sigma_1$ (one may alternatively use $\Sigma_2$).  As noted in \cite{Wall:2012uf} (theorem 3), since the bulk satisfies NCC and the boundary of the future contains only null geodesics without conjugate points, the focusing theorem \cite{Hawking:1969sw} guarantees the representative to have no more area than $x(A)$. And since $\partial \Sigma$ is fixed to be $C_\partial$, the representative must have the same anchor set as $x(A)$.  If $\Sigma$ is not entirely to the future of $\Sigma_1$, one may similarly use e.g. the union of the boundary of the future of $\Sigma_1$ and the boundary of the past of $\Sigma_2$ (or alternatively other combinations of the futures and pasts of $\Sigma_{1,2}$).   And the representative on $\Sigma_{M_R (A,C_\partial)}$ must have area at least as great as $M_R (A,C_\partial)$ since the latter surface is minimal on $\Sigma_{M_R (A,C_\partial)}$. Thus $\operatorname { Area }[M_R(A, C_\partial)] \leq \operatorname { Area }[\tilde x_\Sigma (A)] \leq \operatorname { Area }[x(A)]$, and $M_R(A, C_\partial)$ is a least-area extremal surface.

\subsection{Existence of HRT surfaces in standard charged and rotating black holes}
\label{sec:existence}

The above result will show that HRT surfaces exist in charged or rotating black hole spacetimes.  Let us begin with the AdS-Reissner-Nordstrom (AdS-RN) solution.  The maximal analytic extension is shown in figure \ref{fig:AdSRN}.  However, since the Cauchy horizons are unstable to forming mass-inflation singularities, it is natural to truncate the solution to the unshaded region between the past and future Cauchy horizons\footnote{If one insists on including regions beyond the Cauchy horizons then, as argued in section 6 of \cite{Hubeny:2013gta}, it appears natural to require the homology surface (used in the homology constraint) to be achronal.  This then requires any entangling surface to again lie in our truncated spacetime between the Cauchy horizons.  So for the purposes of entanglement computations there is no harm in our truncation.}.   Given any boundary region $A \subset \partial M$, we may then choose a boundary Cauchy surface $C_\partial \subset \partial M$ with $C_\partial \supset A$ and construct the restricted maximin surface $M_R (A,C_\partial)$ and the associated bulk Cauchy surface $\Sigma_{M_R (A,C_\partial)}$. For $C_\partial$ to be a full Cauchy surface it must include pieces on both boundaries even if $A$ is contained in a single boundary.

Now, by definition, $M$ includes only finite boundary times.  Since $\Sigma_{M_R (A,C_\partial)}$ is achronal (i.e., no two of its points can be connected by a timelike curve) and ends on $C_\partial$, it cannot reach the Cauchy horizon.  Thus $M_R (A,C_\partial)$ lies in the (smooth) interior of the spacetime and the argument of section \ref{sec:equiv} shows that $M_R (A,C_\partial)$ is also an HRT surface for AdS-RN (truncated at the Cauchy horizons).

Furthermore, it is clear that the same conclusion holds for any AdS-hyperbolic spacetime satisfying i) NCC, ii) the generic condition, and for which iii) all bulk Cauchy surfaces $\Sigma$ anchored on boundary Cauchy surfaces $C_\partial$ meet future or past boundaries only at Kasner-like singularities.  We may then use the analysis of Kasner-like singularities in \cite{Wall:2012uf} to argue as above.  In particular, this is true of the perturbed AdS-RN spacetimes with mass-inflation singularities shown in figure \ref{fig:massinf}.  And it continues to hold when rotation is added to the black holes, again truncating the spacetime at Cauchy horizons and/or mass-inflation singularities.  Furthermore, strong subadditivity follows precisely as in \cite{Wall:2012uf}.

\section{Discussion}
\label{sec:discussion}
We have used restricted maximin surfaces to show the existence of HRT surfaces in a broad class of spacetimes including standard black holes with mass-inflation singularities.  In such cases, it also follows that HRT areas satisfy strong subadditivity.  The above class of solutions is believed to be generic in the class of charged and rotating black holes \cite{Poisson:1989zz,Ori:1991zz,Ori:1992zz,Burko:1997zy,Burko:1997fc,
Dafermos:2003wr,Bhattacharjee:2016zof}.  

As explained in the introduction, our result forbids such spacetimes from displaying the HRT-pathologies found in the examples of \cite{Fischetti:2014uxa}.  Taken together with the Lewkowycz-Maldacena \cite{Lewkowycz:2013nqa} and Dong-Lewkowycz-Rangamani \cite{Dong:2016hjy}
derivations, this strongly suggests that these HRT surfaces correctly compute the associated entropies of the dual CFT state.\footnote{These arguments presuppose that the bulk geometry gives the dominant bulk saddle to an appropriate path integral defining a dual CFT state. Although the   the class of Lorentzian black hole spacetimes satisfying this criterion have not been fully characterized, it is clear from e.g.  \cite{Maldacena:2001kr,Balasubramanian:2014hda,Marolf:2015vma,Maxfield:2016mwh,Marolf:2017kvq,Marolf:2018ldl,Botta-Cantcheff:2019apr} that it includes many familiar open sets in the space of solutions.}

\begin{figure}
\begin{center}
  \includegraphics[width=0.4\textwidth]{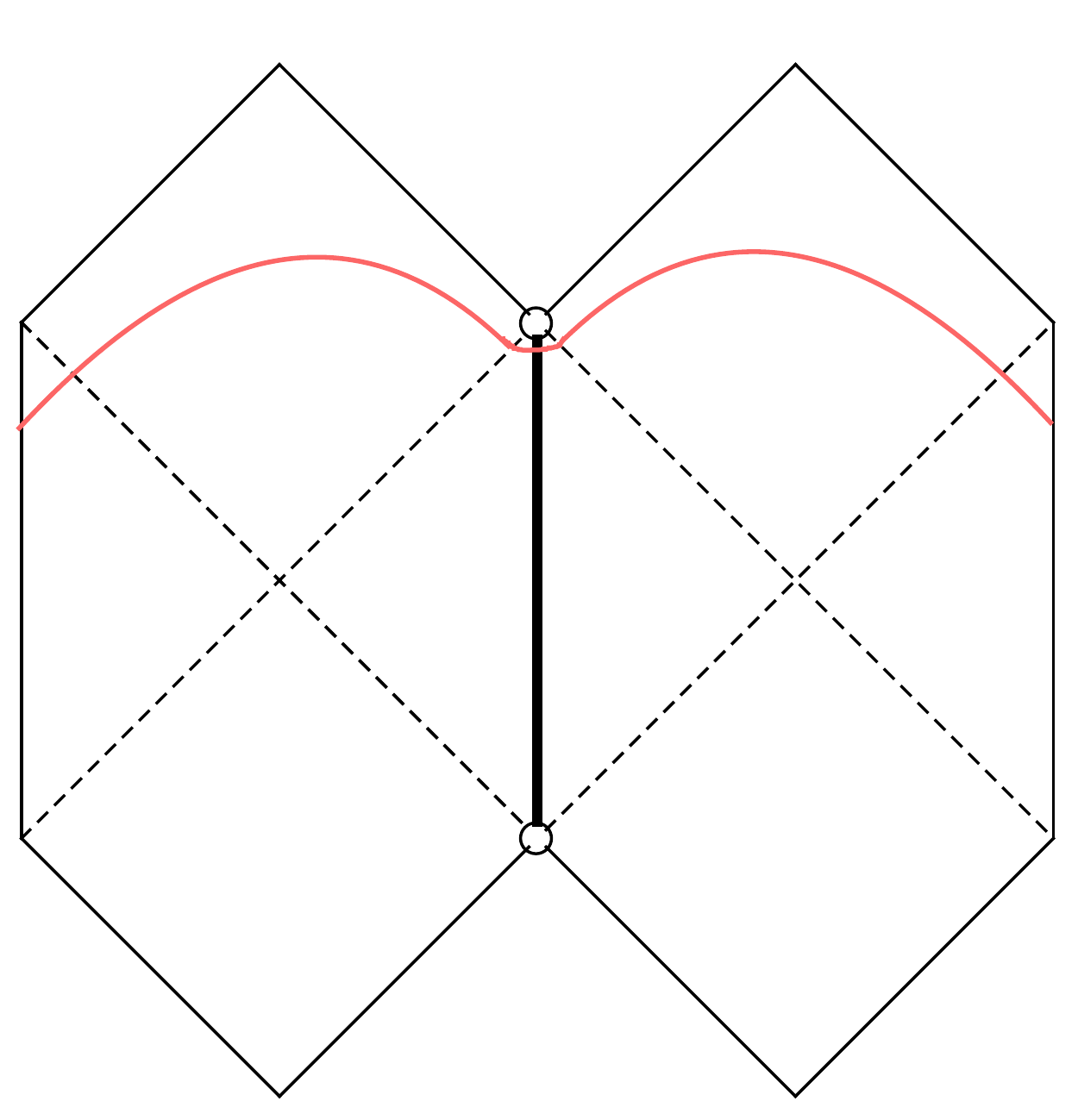}
  \caption{Time-independent charged wormhole which is constructed by sewing two AdS-RN spacetimes together along a domain wall (thick line). Due to the internal infinities (small circles), the Cauchy horizons are union of the Cauchy horizons of the two AdS-RN spacetimes. Limits of Cauchy surfaces like the one shown (red) can reach such horizons.}
  \label{fig:TimeIndep}
 \end{center}
\end{figure}

While our requirements are expected to be satisfied generically, one can nevertheless imagine spacetimes where they fail.  Indeed, generalizing the time-independent wormholes of \cite{Fu:2016xaa} to include electric charge immediately yields solutions of the sort shown in figure \ref{fig:TimeIndep} in which (limits of) boundary-anchored bulk Cauchy surfaces {\it can} reach the bulk Cauchy horizons.    For this particular spacetime one may nevertheless use the fact that the right-most and left-most wedges are identical to those of AdS-RN to show that, for any $A$, there is a (perhaps disconnected) extremal surface anchored to $\partial A$ that is entirely contained in the union of these wedges.  Thus HRT surfaces again exist for this spacetime, but it remains to argue that smaller such surfaces have not been lost to the future and past boundaries.  Other interesting spacetimes may remain to be investigated as well.

\section*{Acknowledgements} We would like to thank Sean Jason Weinberg, Dan Harlow, and Netta Engelhardt for related conversations.  DM was supported in part by the U.S. National Science Foundation under grant PHY 1801805 and by the University of California.   ZW was supported by a Regents Fellowship from the University of California.  AW was supported by the Simons Foundation ``It from Qubit'' grant, AFOSR grant number FA9550-16-1-0082, the John Templeton Foundation under Grant ID\# 60933, the Kavli Institute for Theoretical Physics (supported in part by NSF grant PHY-1748958), and the Stanford Institute for Theoretical Physics.  The opinions expressed are those of the authors and do not necessarily reflect the views of funding agencies.

\bibliographystyle{jhep}
	\cleardoublepage

\renewcommand*{\bibname}{References}

\bibliography{maximin2}

\end{document}